\documentclass{acm_proc_article-sp}

\newcommand{\mycomment}[1]{}
\newcommand{\ie}[0]{{\it i.e.,\ }}
\newcommand{\viz}[0]{{\it viz.,\ }}

\newcommand{\etal}[0]{{\it  et al.\ }}

\makeatletter
\renewcommand*{\p@section}{\S\,}
\renewcommand*{\p@subsection}{\S\,}
\renewcommand*{\p@subsubsection}{\S\,}
\makeatother

\usepackage[normalem]{ulem}

\begin{document}

\title{Unsupervised Anomaly-based Malware Detection using Hardware Features}

\numberofauthors{1} 
\author{
\alignauthor
Adrian Tang \hspace{0.2in} Simha Sethumadhavan \hspace{0.2in} Salvatore Stolfo\\
	   \confname{\hspace{1in}}\\
       \affaddr{Department of Computer Science}\\
       \affaddr{Columbia University}\\
       \affaddr{New York, NY, USA}\\
       \email{\texttt{\{atang, simha, sal\}@cs.columbia.edu}}
}

\date{}
\maketitle

\thispagestyle{empty}

\begin{abstract} 

Recent works have shown promise in using microarchitectural execution
patterns to detect malware programs. These detectors belong to a
class of detectors known as signature-based detectors as they
catch malware by comparing a program's execution pattern (signature)
to execution patterns of \emph{known} malware programs.  In this
work, we propose a new class of detectors --- anomaly-based hardware
malware detectors --- that do not require signatures for malware
detection, and thus can catch a wider range of malware including
potentially novel ones.  We use unsupervised machine learning to
build profiles of normal program execution based on data from
performance counters, and use these profiles to detect significant
deviations in program behavior that occur as a result of malware
exploitation. We show that real-world exploitation of popular
programs such as IE and Adobe PDF Reader on a Windows/x86 platform
can be detected with nearly perfect certainty. We also examine the
limits and challenges in implementing this approach in face of a
sophisticated adversary attempting to evade anomaly-based detection.
The proposed detector is complementary to previously proposed
signature-based detectors and can be used together to improve
security.

\end{abstract}

\section{Introduction}

Malware infections have plagued organizations and users for years,
and are growing stealthier and increasing in number by the day.  In
response to this trend, defenders have created commercial anti-virus
protections, and are actively researching better ways to detect
malware.  An emerging and promising approach to detect malware is
to build malware detection systems in
hardware~\cite{Demme:2013:FOM:2485922.2485970}. The idea is to use
information easily available in hardware (typically through performance
counters) to detect malware. It has been argued that hardware malware
schemes are desirable for two reasons: first, unlike software malware
solutions that aim to protect vulnerable software with equally
vulnerable software~\footnote{Software antivirus (AV) systems roughly
have the same bug defect density as regular software.}, hardware
systems protect vulnerable software with robust hardware implementations
that have lower bug defect density because of their simplicity.
Second, while a motivated adversary can evade either defense, evasion
is harder in a system that utilizes hardware features. The intuition is that
the attacker does not have the same degree of control over low-level
hardware execution features as she has with software features. For
instance, it is easier to change system calls or file names than
modify cache hit rates and branch predictor rates in a really precise
way across a range of time scales while still exploiting the system.

\begin{figure}
  \centering
  \includegraphics[width=\columnwidth]{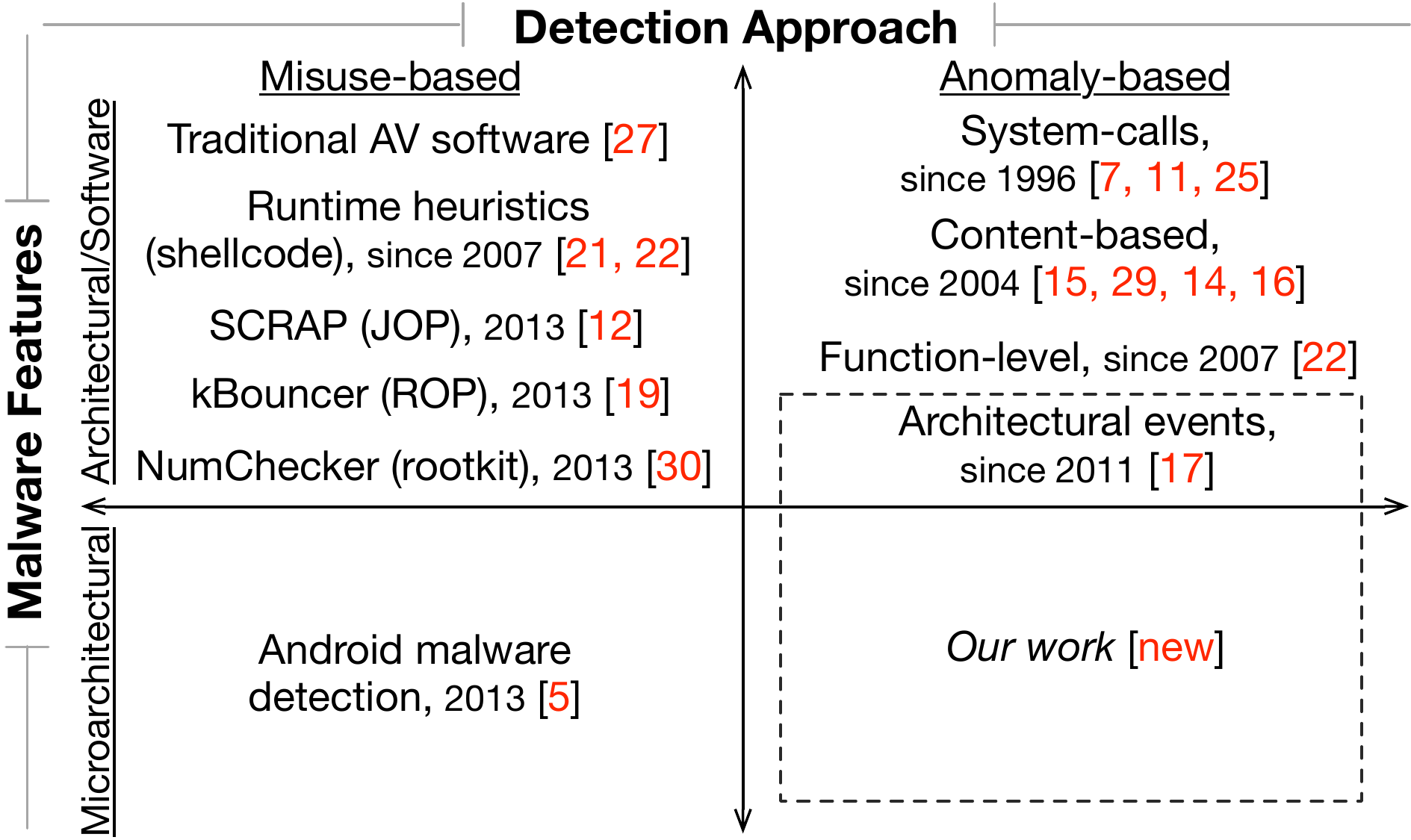}
  \caption{Taxonomy of malware detection approaches and some example works.}
  \label{fig:taxonomy}
\end{figure}

In this paper we introduce a new class of malware detectors known
as hardware anomaly-based detectors. All existing malware detection
techniques, software or hardware, can be classified along two
dimensions: \textit{detection approach} and the \textit{malware
features} they target, as presented in Figure~\ref{fig:taxonomy}.
Detection approaches are traditionally categorized into
misuse-based and anomaly-based detection. Misuse detection attempts
to flag malware based on pre-identified execution signatures or
attack patterns.  It can be highly accurate against known attacks,
but is extremely susceptible to attacks with slight modifications
deviating from the signatures.  On the other hand, anomaly-based
detection characterizes baseline models of a state of normalcy and
identifies attacks based on deviations from the models. Other than
being able to target a wide range of attacks, it can potentially
identify novel ones. There are a range of features that can be used
for detection: until 2013, features used for malware detection were
software features such as system call signatures and patterns, or
network traffic. Since then, features available in hardware including
microarchitectural features have been used for malware detection.
As shown in Figure~\ref{fig:taxonomy}, we examine for the first
time, the feasibility and limits of performing anomaly-based malware
detection using low-level architectural and microarchitectural
features available from hardware performance counters (HPCs).

A typical malware infection can be understood as a two-stage process,
exploitation and take-over. In the exploitation stage, an adversary
exercises a bug in the victim program to hijack control of the
victim program execution. Exploitation is then followed by more
elaborate procedures to download and install a payload such as a
keylogger.  Prior work on hardware-level malware detection such as \cite{Demme:2013:FOM:2485922.2485970} has focused on flagging Android malicious apps by detecting payloads. 
Detecting malware during exploitation
not only gives more lead time for mitigations but can also act as
an early threat predictor to improve the accuracy of subsequent
signature-based detection of payloads.

The key intuition for anomaly-based detection stems from the
observation that the malware, during the exploitation stage, alters
the original program flow to execute peculiar non-native code in
the context of the victim program. Such unusual code execution will
cause perturbations to dynamic execution characteristics of the
program, and if these perturbations are observable they can form
the basis for the detection of malware exploits. Since exploits
manipulate execution flow within the victim program, the signature-based
detection paradigm is not appropriate for detecting exploitation.
For instance, a signature-based detector will likely correctly
report that IE is executing even when it is infected with malware
because the malware executes alongside IE.

\nocite{szor2005art, polychronakis2007emulation, polychronakis2010comprehensive, kayaalp2013scrap, Wang:2013:NDK:2463209.2488831}

In this paper, we model the characteristics of common vulnerable
programs such as Internet Explorer 8 and Adobe PDF Reader 9 -- two
of the most attacked programs, -- and investigate if and to what
degree malware code execution causes observable perturbations to
these characteristics.  In this anomaly-based detection approach,
intuitively one might expect the deviations caused by exploits to
be fairly small and unreliable, especially in vulnerable programs
with very varied use such as the ones we used. This intuition is
validated in our measurements. On a production Windows machine
running on Intel x86 chips, our experiments indicate that distributions
of execution measurements from the hardware performance counters
are positively skewed, with many values being clustered near zero.
This implies minute deviations caused by the exploit code cannot
be effectively discerned directly.

However, we show that this problem of identifying deviations from
the heavily skewed distributions can be alleviated.  We show that
by using the power transform to amplify small differences, together with
temporal aggregation of multiple samples, we are able to set apart
the execution of the exploit within the context of the larger program
execution.  Further, in a series of experiments, we systematically
evaluate the detection efficacy of the models over a range of
operational factors, events selected for modeling and sampling
granularity. For IE exploitation, we are able to identify 100\% of the exploitation epochs with 1.1\% false positives. Since exploitation typically occurs across nearly 20 epochs, even with a slightly lower true positive rate with high probability we catch exploitations.
 These results are achieved at
a sampling overhead of 1.5\% slowdown using sampling granularity
of 512K instructions epochs.

Further, we examine resilience of our detection technique to evasion
strategies of a sophisticated adversary.  Specifically we model
attackers who conduct \textit{mimicry} attacks by crafting malware
to exhibit event characteristics that resemble normal code execution
to evade our anomaly detection models.  With generously optimistic
assumptions about attacker and system capabilities,  we demonstrate
that the models are susceptible to the mimicry attack.  In a worst
case scenario, the detection performance deteriorates by up to
6.5\%.  Due to this limitation we observe that anomaly detectors
cannot be the only defensive solution but can be valuable as part
of an ensemble of predictors that can include signature-based
predictors.

The rest of the paper is organized as follows. We provide a background 
on modern malware exploits in \ref{sec:background}. We detail our 
experimental setup in \ref{sec:experiment}. We present our approach in 
building models for the study in \ref{sec:models}, and describe
the experimental results in \ref{sec:results}. \ref{sec:evasion} examines  
evasion strategies of an adaptive adversary and the impact on
detection performance. In \ref{sec:arch}, we discuss architectural 
enhancements that will facilitate better malware detection 
at the hardware level. \ref{sec:related} discusses related work, and 
we conclude in \ref{sec:conclusion}.

\section{Background}
\label{sec:background}

\begin{figure}
  \centering
  \includegraphics[width=\columnwidth]{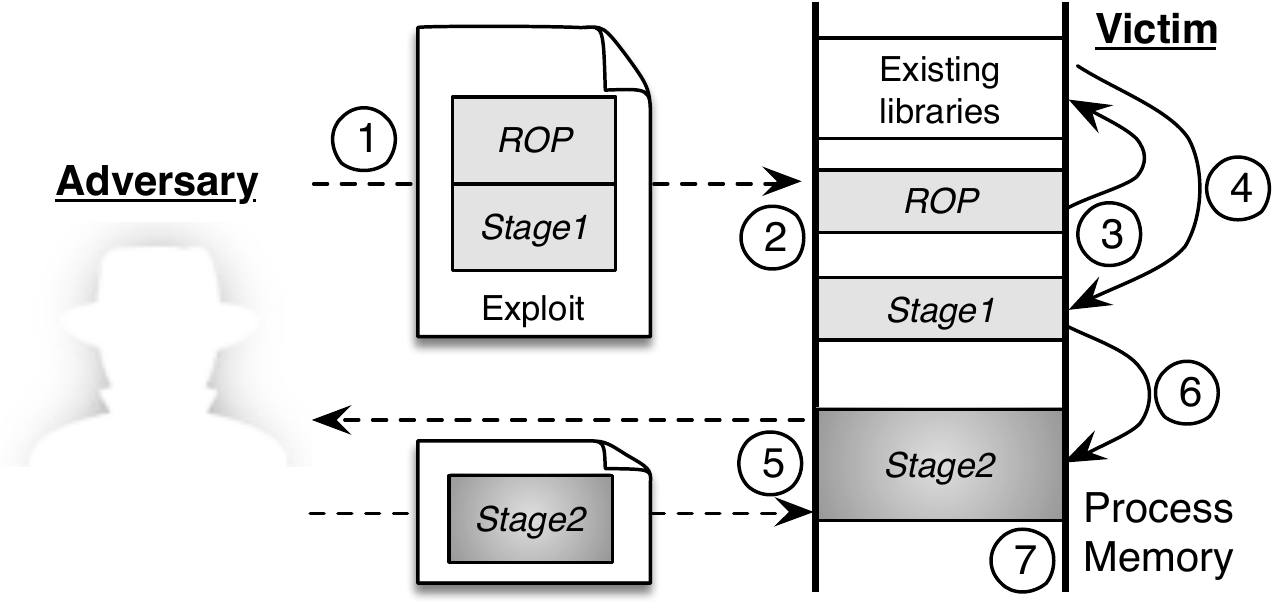}
  \caption{Multi-stage exploit process.}
  \label{fig:multi_stage_attack}
\end{figure}

Figure~\ref{fig:multi_stage_attack} shows a typical multi-stage
malware infection process that results in a system compromise. The
necessity for its multi-stage nature will become clear as
we explain the exploit process in this section.

\textbf{Triggering the vulnerability} \space\space First the adversary
crafts and delivers the exploit to the victim to target a specific
vulnerability known to the adversary (Step
\raisebox{.1pt}{\textcircled{\raisebox{-.5pt} {1}}}). The vulnerability
is typically a memory corruption bug; the exploit is typically
sent to a victim from a webpage or a document attachment from an email.
When the victim accesses
the exploit, two exploit subprograms, commonly known as the
\textit{ROP} and \textit{Stage1} ``shellcodes'' load into the memory
of the vulnerable program (Step
\raisebox{.1pt}{\textcircled{\raisebox{-.5pt} {2}}}). The exploit
then uses the vulnerability to transfer control to the \textit{ROP}
shellcode (Step \raisebox{.1pt}{\textcircled{\raisebox{-.5pt} {3}}}).

\textbf{Code Reuse Shellcode (\textit{ROP})} \space\space To
prevent untrusted data being executed
as code, modern processors provide Data Execution Prevention (DEP)
to prevent code from being run from data pages.  A problem is that
DEP can be toggled by the program itself. This feature is necessary
to support JIT compilation.  So, to circumvent DEP, the \textit{ROP}-stage
shellcode reuses instructions in the original program
binary -- hence the name Code Reuse Shellcode -- to craft a call
to the function that disables DEP for the data page containing the
next \textit{Stage1} shellcode. The ROP ShellCode then redirects
execution to the next stage.  (Step
\raisebox{.1pt}{\textcircled{\raisebox{-.5pt} {4}}}) \cite{Corelan:2011,
pappastransparent}.

\textbf{\textit{Stage1} Shellcode} \space\space This shellcode is
typically a relatively small -- from a few bytes to about 300
bytes~\footnote{As observed at \url{http://exploit-db.com}} -- code
stub with exactly one purpose: to download a larger (evil) payload
which can be run more freely. To maintain stealth, it downloads the
payload in memory (Step \raisebox{.1pt}{\textcircled{\raisebox{-.5pt}
{5}}}). 

\textbf{\textit{Stage2} Payload} \space\space The payload is the
final piece of code that the adversary wants to execute on the user
machine to perform a specific malicious task.  The range of
functionality of this payload, commonly a backdoor, keylogger, or
reconnaissance program, is in fact unlimited.  After the payload is
downloaded, the \textit{Stage1} shellcode runs that payload as an
executable using reflective DLL injection (Step
\raisebox{.1pt}{\textcircled{\raisebox{-.5pt} {6}}}), a stealthy
library injection technique that does not require any physical files
\cite{Fewer:2008}.  By this time, the victim system is fully
compromised (Step \raisebox{.1pt}{\textcircled{\raisebox{-.5pt}
{7}}}).

The \textit{Stage1} shellcode and \textit{Stage2} payload are
different in terms of size, design and function, primarily due to
the operational constraints on the \textit{Stage1} shellcode. When
delivering the initial shellcode in the exploit, exploit writers
typically try to use as little memory as possible to ensure that
the program does not unintentionally overwrite their exploit code
in memory. To have a good probability for success this code needs
to be small and fast, and thus is written in assembly in very
restrictive position-independent memory addressing style. These
constraints limit the attackers ability to write very large shellcodes.
In contrast, the \textit{Stage2} payload does not have all these
constraints and can be developed like any regular program.  This
is similar to how OSes use small assembly routines to bootstrap and
then switch to compiled code.

The strategy and structure described above is representative of a
large number of malware especially those created with fairly recent
web exploit kits~\cite{TM:2012}.  These malware exploits execute
completely from memory and in the process context of the host victim
program.  Further, they maintain disk and process stealth by ensuring
no files are written to disk and no new processes are created, and
thus easily evade most file based malware detection techniques.

\section{Experimental Setup}
\label{sec:experiment}
Do the execution of different shellcode stages exhibit observable deviations from the baseline performance characteristics of the user programs? Can we use these deviations, if any, to detect a malware exploit as early as possible in the infection process? To address these questions, we conduct several feasibility experiments, by building baseline per-program models using machine learning classifiers and examining their detection efficacy over a range of operational factors. Here, we describe various aspects of our experimental setup and detail how we collect and label the performance measurements that are attributed to different stages of malware exploits.

\subsection{Exploits}

Unlike SPEC~\cite{Henning:2006:SCB:1186736.1186737}, there are no standard exploit benchmarks.  We created
our own exploits for common vulnerable programs from publicly
available information.  We use exploits that target the security
vulnerabilities \textit{CVE-2012-4792}, \textit{CVE-2012-1535} and
\textit{CVE-2010-2883} on IE 8 and the web plug-ins, \ie  Adobe
Flash 11.3.300.257 and Adobe Reader 9.3.4 respectively.  We generated
the exploits using a widely-used penetration testing tool
\textit{Metasploit}~\footnote{\url{http://www.metasploit.com/}}. We use \textit{Metasploit}
because the exploitation techniques it employs in the exploits are
representative of multi-stage nature of real-world exploits.

With \textit{Metasploit}, besides targeting different vulnerabilities
using different ROP shellcode, we also vary both the \textit{Stage1}
shellcode and the \textit{Stage2} final payload used in the exploits.
The variability in the generated exploits is summarized in Table
\ref{tbl:malcode_variability}. For the \textit{ROP} shellcode stage,
the relevant library files, from where the shellcode is derived,
are listed.

\subsection{Measurement Infrastructure}
\label{sec:m_infra}

Since most real-world exploits run on Windows and PDF readers,
and none of the architectural simulators can run programs of this
scale, we use measurements from production machines.  We develop
a Windows driver to configure the performance monitoring unit on
Intel i7 2.7GHz IvyBridge Processor to interrupt once every \textit{N}
instructions and collect the event counts from the HPCs. We also 
record the Process ID (PID) of the currently executing program so 
that we can filter the measurements based on processes.  

We collected the measurements from a VMware Virtual Machine (VM)
environment, installed with Windows XP SP3 and running a single-core
with 512MB of memory. With the virtualized HPCs in the VM, this
processor enables the counting of two fixed events (clock cycles,
instruction retired) and up to a limit of four events simultaneously.
We configured the HPCs to update the event counts only in the user
mode.

To ensure experiment fidelity for the initial study, the measurements
from the memory buffer are read and transferred via TCP network
sockets to a \textit{recorder} that is deployed in another VM. This
\textit{recorder} saves the stream of measurements in a local file
that is used for our analysis.

\subsection{Sampling Granularity}
We experimented with various sampling interval of \textit{N}
instructions. We chose to begin the investigation with a sampling
rate of every 512,000 instructions since it provides a reasonable
amount of measurements without incurring too much overhead (See 
\ref{sec:sampling_overhead} for an evaluation of the sampling
overhead). Each sample consists of the event counts from one sampling
time epoch, along with the identifying PID and exploit stage label.

\subsection{Collection of Clean and Infected \\Measurements} \label{sec:measurements}

To obtain clean exploit-free measurements for IE 8, we randomly
browsed websites that use different popular web plugins available
on IE \viz Flash, Java, PDF, Silverlight, and  Windows Media Player
extensions.  We visited the top 20 websites from Alexa and included several other websites that to widen the coverage of the use of the various plug-ins. Within the browser, we introduced variability by randomizing the order in which the websites are loaded across runs; likewise we accessed websites by clicking links randomly and manually on the webpages. The dynamic content on the websites also perturbs the browser caches. We used a maximum of two concurrent tabs. In addition, we simulated plug-in download and installation functions.

For Adobe PDF
measurements, we downloaded 800 random PDFs from the web, reserving half of them randomly for training and the other half for testing.

To gather infected measurements, we browse pages with our PDF exploits
with the same IE browser that uses the PDF plug-in. We use Metasploit to
generate these PDF exploits and ensure that both the clean
and unclean PDFs had the same distribution of file types, for
instance, same amount of Javascript.

We stop gathering infected measurements when we see creation of a
new process. Usually the target process becomes unstable due to the
corrupted memory state, and the malicious code typically ``migrates''
itself to another new or existing process to ensure persistence
after the execution of the \textit{Stage2} payload. This is an 
indication that the infection is complete.

We use the same input sets for different measurements, and between
each run of the exploit, we revert the VM environment to ensure the
samples collected from the next run is not contaminated from the
previous run.

\begin{table}
\centering
\footnotesize
\begin{tabular}{|l|l|} \hline
\textbf{Stage} & \textbf{Variation} \\ \hline \hline
ROP & \texttt{msvcrt.dll}, \texttt{icucnv36.dll}, \texttt{flash32.ocx} \\ \hline
Stage1 & reverse\_tcp, reverse\_http, bind\_tcp \\ \hline
Stage2 & meterpreter, vncinject, command\_shell \\ \hline
\end{tabular}
\caption{Variability in exploit code.}
\label{tbl:malcode_variability}
\end{table}

\subsection{Bias Mitigation}

While there are factors that may affect the results of our measurements,
we took care to mitigate some possible biases in our data by ensuring
the following during the measurement collection.

\textbf{Between-run contamination} \space\space After executing
each exploit and collecting the measurements, we restore the VM to
the state before the exploit is exercised. This ensures the
measurements collected are independent across training and testing
sets, and across different \textit{clean} and exploit runs.

\textbf{Exploitation bias} \space\space Loading the exploits in the
program in only one way may bias the sampled measurements. To reduce
this bias, we collected the measurements while loading the exploit
in different ways: (1) We launch the program and load the URL link of the
generated exploit page. (2) With an already running program instance,
we load the exploit page.  (3) We save the exploit URL in a shortcut
file and launch the link shortcut with the program.

\textbf{Network conditions} \space\space The VM environment is
connected to the Internet. To ensure that the different network
latencies do not confound the measurements, we configure the VM
environment to connect to an internally-configured
\textit{Squid}~\footnote{\url{http://www.squid-cache.org/}} proxy 
and throttle the network bandwidth from 0.5 to 5Mbps using \textit{Squid} delay pools. We vary 
the bandwidth limits while collecting measurements for both the exploit 
code execution and clean runs.

\section{Building Models}
\label{sec:models}

To use HPC measurements for anomaly-based detection
of malware exploits, we need to build classification models to
describe the baseline characteristics for each program we
protect. These program characteristics are relatively rich in
information and, given numerous programs, manually building the 
models is nearly impossible.  Instead we rely on unsupervised machine
learning techniques to dynamically learn possible hidden structure
in these data. We then use this hidden structure -- aka model --
to detect deviations during exploitation.


We rely on a class of \textit{unsupervised} one-class machine
learning techniques for model building. The one-class approach is
very useful because the classifier can be trained \emph{solely}
with measurements taken from a clean environment. This removes  
the need to gather measurements affected by exploit code, which
is hard to implement and gather in practice. Specifically, we model the characteristics with the one-class Support Vector Machine (oc-SVM) classifier that uses the non-linear Radial Basis Function (RBF) kernel
\cite{Scholkopf:2001:ESH:1119748.1119749}. In this study, the
collection of the labeled measurements is purely for evaluating how
effective models are in distinguishing measurements taken in the
presence of malware code execution.

\begin{table}
\centering
\footnotesize
\begin{tabular}{|l|l|} \hline
\textbf{Name} & \textbf{Event Description}\\ \hline \hline
	\multicolumn{2}{|c|}{\textbf{Architectural Events}}\\ \hline
\textsc{Load} & Load instructions retired\\ \hline
\textsc{Store} & Store instructions retired\\ \hline
\textsc{Arith} & Arithmetic instructions retired\\ \hline
\textsc{Br} & Branch instructions retired \\ \hline
\textsc{Call} & All near call instructions retired\\ \hline
\textsc{Call\_D} & Direct near call instructions retired\\ \hline
\textsc{Call\_ID} & Indirect near call instructions retired\\ \hline
\textsc{Ret} & Near return instructions retired \\ \hline \hline
	\multicolumn{2}{|c|}{\textbf{Microarchitectural Events}}\\ \hline
\textsc{Llc} & Last level cache references\\ \hline
\textsc{Mis\_Llc} & Last level cache misses\\ \hline
\textsc{Misp\_Br} & Mispredicted branch instructions\\ \hline
\textsc{Misp\_Ret} & Mispredicted near return instructions\\  \hline
\textsc{Misp\_Call} & Mispredicted near call instructions\\ \hline
\textsc{Misp\_Br\_C} & Mispredicted conditional branch \\ \hline
\textsc{Mis\_Icache} & I-Cache misses\\ \hline
\textsc{Mis\_Itlb} & I-TLB misses\\ \hline
\textsc{Mis\_Dtlbl} & D-TLB load misses\\ \hline
\textsc{Mis\_Dtlbs} & D-TLB store misses\\ \hline
\textsc{Stlb\_Hit} & Shared-TLB hits after i-TLB misses\\ \hline
\textsc{\%Mis\_Llc}\footnotemark[5] & \% of last level cache misses \\ \hline
\textsc{\%Misp\_Br}\footnotemark[5] & \% of mispredicted branches\\ \hline
\textsc{\%Misp\_Ret}\footnotemark[5] & \% of mispredicted near RET instructions\\ \hline
\end{tabular}
\caption{Shortlisted candidate events to be monitored.}
\label{tbl:eventtypes}
\end{table}
	\footnotetext[5]{These \textit{derived} events are not directly measured, but are computed using two events monitored and measured by the HPCs. For example, \textsc{\%Misp\_Br} is computed as $\textsc{Misp\_Br}/\textsc{Br}$.}

\subsection{Feature Selection}

While the Intel processor we use for our measurements permits hundreds
of events to be monitored using HPCs, not all of them 
are equally useful in characterizing the execution of programs. We 
examine most of the events investigated in previous program 
characterization works \cite{Shen:2008:HCD:1346281.1346306, Hoste:2012}, 
and various other events informed by our understanding on the malware behavior. 
Out of the hundreds of possible events that can be 
monitored, we shortlisted  19 events for this study in Table 
\ref{tbl:eventtypes}. We further differentiate between the \textit{Architectural} 
events that give an indication of the execution mix of instructions in any running program, 
and the \textit{Microarchitectural} ones that are dependent on
the specific hardware makeup of a system.

\textbf{Events with higher discriminative powers} \space\space The 
processor is limited to monitoring only up to 4 events at any given time. 
Even with the smaller list of shortlisted events, we have to select 
only a subset of events, aka features, that can most effectively 
differentiate clean execution from infected execution. Since we 
have at our disposal labeled measurements, we use the
Fisher Score (\emph{F-Score}) to provide a quantitative measure of
the how effective a feature can discriminate between measurements 
in clean executions from those in infected executions. In general, the 
F-Score is a widely-used feature selection metric that measures the 
discriminative power of features \cite{2007:RDP:1325591.1325635}. 
A feature with better discriminative power would have a larger separation between the means and standard
deviations for samples from different classes. The F-Score gives a
measure of this degree of separation. The larger the F-Score, the
more discriminative power the feature is likely to have. However,
a limitation to using the F-Score is that it does not account for
mutual information/dependence between features, but it helps guide
our selection of a subset of ``more useful'' features.

Since we are attempting to differentiate samples with malicious
code execution from those without, we compute the corresponding F-Scores 
for each event. We compute the F-Scores for the different stages
of malware code execution for each event and reduce the
shortlisted events to the 7 top-ranked events for each of the two
categories, as well as for the two categories combined, in Table 
\ref{tbl:top_ranked_fscores}. Each row consists of the top-ranked
events for an event category and the exploit stage.

We further select the top 4 events from each row to form 9 candidate 
event sets that we will use to build 
the baseline characteristics models of the IE browser.
Each model constructed with one set of events can then be evaluated 
for its effectiveness in the detection of various stages of malware 
code execution. For brevity, we assign a label (such as \textit{A-0} 
and \textit{AM-2}) to each set of 4 events in Table 
\ref{tbl:top_ranked_fscores} and refer to each model based on this
\textit{set label}. We note that the derived events such as 
\textsc{\%Misp\_Br} are listed in the table solely for comparison. 
Computing them requires monitoring two events and reduces the number
of features used in the models. Via experimentation, we find that 
using them in the models does not increase the efficacy of the models. 
So, we exclude them from the event sets.

\textbf{Feature Extraction} \space\space Each sample consists of 
simultaneous measurements of all the four event counts in one time epoch. 
We convert the measurements in each sample to the vector subspace, so 
that each classification vector is represented as as a four-feature 
vector. Each vector, using this feature extraction method, represents 
the measurements taken at the smallest time-slice for that sampling 
granularity. These features will be used to build \textit{non-temporal} models.

Since we observe that malware shellcode typically runs over several time 
epochs, there may exist temporal relationships in the measurements 
that can be exploited. To model any potential temporal information, 
we extend the dimensionality of each sample vector by grouping the 
\textit{N} consecutive samples and combining the measurements of each 
event to form a vector with $4N$ features. We use $N=4$ to create 
sample vectors consisting of 16 features each, so each sample vector 
effectively represents measurements across 4 time epochs. By grouping
samples across several time epochs, we use the synthesis of these 
event measurements to build \textit{temporal} models.

With the 
granularity at which we sample the measurements, the execution of 
the ROP shellcode occurs within the span of just one sample. Since we 
are creating vectors with a number of samples as a group, the ROP 
payload will only contribute to one small portion of a vector sample. 
So we leave out the ROP shellcode for testing using this form of 
feature extraction.

\begin{table*}
\resizebox{\textwidth}{!}{%
\centering

\begin{tabular}{|c|c||c|c|c|c|c|c|c|c|} \hline
\textbf{Exploit} & \textbf{Set} & \multicolumn{8}{c|}{\textbf{Events ranked by F-scores}}\\ \cline{3-10}
\textbf{Stage} & \textbf{Label} & \textbf{1} & \textbf{2} & \textbf{3} & \textbf{4} & \textbf{5} & \textbf{6} & \textbf{7} & \textbf{8} \\ \hline \hline

	\multicolumn{10}{|c|}{\textbf{A}rchitectural Events}\\ \hline

	ROP & A-0 & {\color{RoyalBlue}\textbf{\textsc{Ret}}} & {\color{RoyalBlue}\textbf{\textsc{Call\_D}}} & {\color{RoyalBlue}\textbf{\textsc{Store}}} & {\color{RoyalBlue}\textbf{\textsc{Arith}}} & \textsc{Call} & \textsc{Load} & \textsc{Call\_Id} & \textsc{Br} \\ \hline
	Stage1 & A-1 & {\color{RoyalBlue}\textbf{\textsc{Store}}} & {\color{RoyalBlue}\textbf{\textsc{Load}}} & {\color{RoyalBlue}\textbf{\textsc{Call\_ID}}} & {\color{RoyalBlue}\textbf{\textsc{Ret}}} & \textsc{Call\_D} & \textsc{Call} & \textsc{Arith} & \textsc{Br}\\ \hline
	Stage2 & A-2 & {\color{RoyalBlue}\textbf{\textsc{Store}}} & {\color{RoyalBlue}\textbf{\textsc{Call\_ID}}} & {\color{RoyalBlue}\textbf{\textsc{Ret}}} & {\color{RoyalBlue}\textbf{\textsc{Call\_D}}} & \textsc{Call} & \textsc{Arith} & \textsc{Br} & \textsc{Load} \\ \hline \hline
	
	\multicolumn{10}{|c|}{\textbf{M}icroarchitectural Events}\\ \hline 

	ROP & M-0 & {\color{RoyalBlue}\textbf{\textsc{Misp\_Br\_C}}} & \textsc{\%Misp\_Br} & {\color{RoyalBlue}\textbf{\textsc{Misp\_Br}}} & \textsc{\%Misp\_Ret} & {\color{RoyalBlue}\textbf{\textsc{Mis\_Itlb}}} & {\color{RoyalBlue}\textbf{\textsc{Mis\_Llc}}} & \textsc{Mis\_Dtlbs} & \textsc{Misp\_Call} \\ \hline
	Stage1 & M-1 & {\color{RoyalBlue}\textbf{\textsc{Misp\_Ret}}} & {\color{RoyalBlue}\textbf{\textsc{Misp\_Br\_C}}} & \textsc{\%Misp\_Ret} & \textsc{\%Misp\_Br} & {\color{RoyalBlue}\textbf{\textsc{Mis\_Dtlbs}}} & {\color{RoyalBlue}\textbf{\textsc{Stlb\_Hit}}} & \textsc{Misp\_Br} & \textsc{Mis\_Icache} \\ \hline
	Stage2 & M-2 & {\color{RoyalBlue}\textbf{\textsc{Misp\_Ret}}} & {\color{RoyalBlue}\textbf{\textsc{Stlb\_Hit}}} & {\color{RoyalBlue}\textbf{\textsc{Mis\_Icache}}} & {\color{RoyalBlue}\textbf{\textsc{Mis\_Itlb}}} & \textsc{\%Misp\_Ret} & \textsc{Misp\_Call} & \textsc{Mis\_Llc} & \textsc{Misp\_Br\_C} \\ \hline \hline
	
	\multicolumn{10}{|c|}{Both \textbf{A}rchitectural and \textbf{M}icroarchitectural Events}\\ \hline

	ROP & AM-0 & {\color{RoyalBlue}\textbf{\textsc{Misp\_Br\_C}}} & \textsc{\%Misp\_Br} & {\color{RoyalBlue}\textbf{\textsc{Misp\_Br}}} & \textsc{\%Misp\_Ret} & {\color{RoyalBlue}\textbf{\textsc{Mis\_Itlb}}} & {\color{RoyalBlue}\textbf{\textsc{Ret}}} & \textsc{Mis\_Llc} & \textsc{Mis\_Dtlbs} \\ \hline
	Stage1 & AM-1 & {\color{RoyalBlue}\textbf{\textsc{Store}}} & {\color{RoyalBlue}\textbf{\textsc{Load}}} & {\color{RoyalBlue}\textbf{\textsc{Misp\_Ret}}} & {\color{RoyalBlue}\textbf{\textsc{Call\_ID}}} & \textsc{Ret} & \textsc{Call\_D} & \textsc{Call} & \textsc{Misp\_Br\_C} \\ \hline
	Stage2 & AM-2 & {\color{RoyalBlue}\textbf{\textsc{Store}}} & {\color{RoyalBlue}\textbf{\textsc{Call\_ID}}} & {\color{RoyalBlue}\textbf{\textsc{Misp\_Ret}}} & {\color{RoyalBlue}\textbf{\textsc{Ret}}} & \textsc{Call\_D} & \textsc{Call} & \textsc{Stlb\_Hit} & \textsc{Mis\_Icache} \\ \hline

\end{tabular}}

\caption{Top 8 most discriminative events for different stages of exploit execution (Each event set consists of 4 event names in {\color{RoyalBlue}\textsc{Bold}}. E.g, monitoring event set \textit{A-0} consists of simultaneously monitoring \textsc{Ret}, \textsc{Call\_D}, \textsc{Store} and \textsc{Arith} event counts.)}
\label{tbl:top_ranked_fscores}
\end{table*}

\section{Results}
\label{sec:results}

\subsection{Anomalies Not Directly Detectable}

We first investigate if we can gain any insights into the distribution of the event counts for a clean environment and one attacked by an exploit. Without assuming any prior knowledge of the distributions, we use the box-and-whisker plots\footnotemark[6] of normalized measurements for the different events. These plots provide a visual gauge of the range and variance in the measurements, and an initial indication on how distinguishable the measurements taken with the execution of different stages of malware code are from the \textit{clean} measurements from an exploit-free environment. 

	\footnotetext[6]{The box-and-whisker plot is constructed with the bottom and top of the box representing the first and third quartiles respectively. The {\color{red}red} line in the box is the median. The whiskers extend to 1.5 times the length of the box. Any outliers beyond the whiskers are plotted as blue \emph{{\color{RoyalBlue}+}} ticks.}

\begin{figure}
  \centering
  \includegraphics[width=\columnwidth]{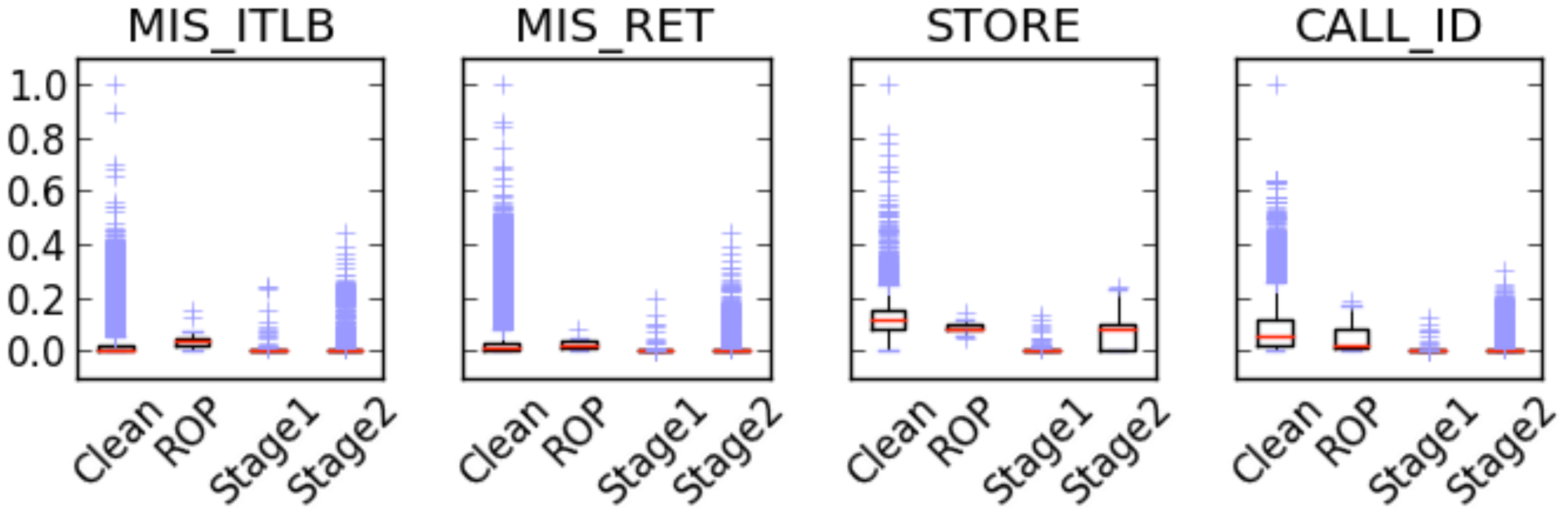}
  \caption{Comparison of distributions of events from \textit{clean} runs versus different malware shellcode stages.}
  \label{fig:univariate_before_sub}
\end{figure}

These distribution comparisons suggest that any event anomalies manifested by malware code execution are not trivially detectable, due to two key observations. (1) Most of the measurement distributions are very positively skewed, with many values clustered near zero. (2) Deviations, if any, from the baseline event characteristics due to the exploit code are not easily discerned. These observations are evident in Figure \ref{fig:univariate_before_sub}, where we present the distribution plots for a few events.

\subsubsection{Power Transform}

To alleviate this challenge, we rely on a rank-preserving power transform on the measurements to positively scale the values. In the field of statistics, the power transform is a family of functions that is commonly applied to data to transform non-normally distributed data to one that has approximately normal distribution. Used in our context, it has the value of magnifying any slight deviations that the malware code execution may have on the baseline characteristics.

For each event type, we find the appropriate power parameter $\lambda$ such that the normalized median is roughly 0.5. We maintain and use this parameter $\lambda_i$ for each event \textit{i} to scale all its corresponding measurements throughout the experiment. Each normalized and scaled event measurement for event \textit{i}, $normalized_i$, is transformed from the raw value $raw_i$ as follows: $normalized_i = {(\frac{raw_i - min_i}{max_i})}^{{\lambda}_i}$, where the $min_i$ and $max_i$ are the minimum and maximum values for this event.

\begin{figure*}
  \centering
  \includegraphics[width=\textwidth]{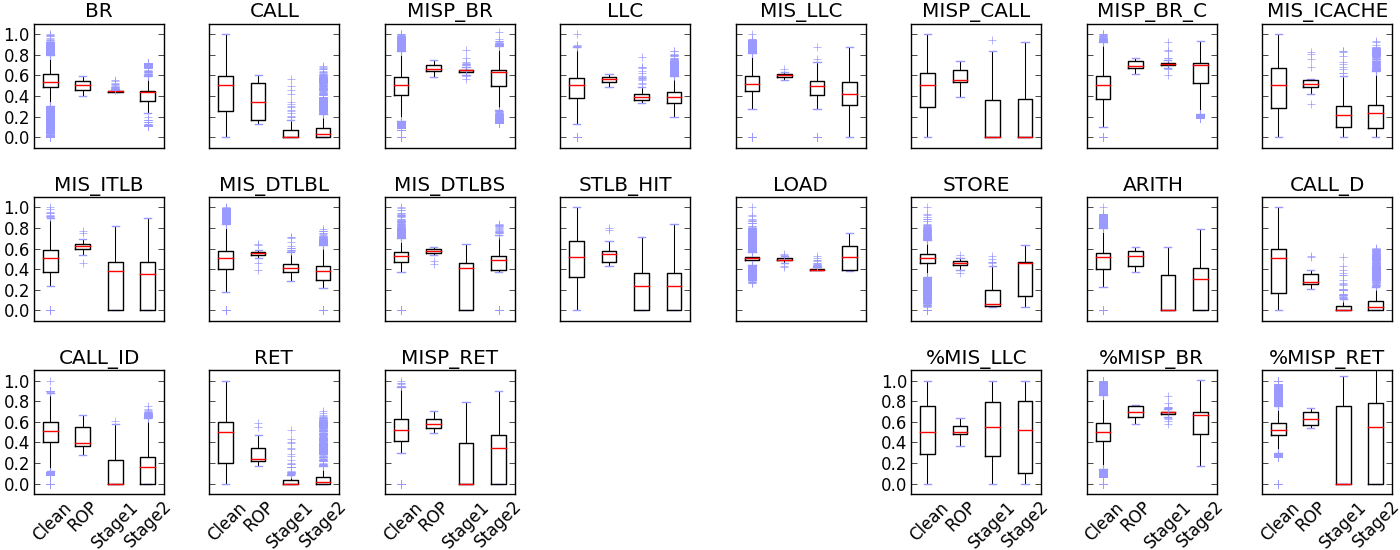}
  \caption{Distribution comparison of all the events (\textit{after} power transform), with more discernible deviations.}
  \label{fig:univariate_after}
\end{figure*}

Using this power transform, we plot the distributions of all the events, in Figure \ref{fig:univariate_after}. Now we observe varying deviations from baseline characteristics due to different stages of malware code execution for various event types. Some events (such as \textsc{Misp\_Ret} and \textsc{Store}) show relatively larger deviations, especially for the \textit{Stage1} exploit shellcode. These events likely possess greater discriminative power in indicating the presence of malware code execution. Clearly, there are also certain events that are visually correlated. The \textsc{Ret} and \textsc{Call} exhibit similar distributions. We can also observe strong correlation between those computed events (such as \textsc{\%Misp\_Br}) and their constituent events (such as \textsc{Misp\_Br}).

\subsection{Evaluation Metrics for Models}

To visualize the classification performance of the models, we construct the \emph{Receiver Operating Characteristic} (ROC) curves which plot the percentage of truely identified malicious samples (True positive rate) against the percentage of \textit{clean} samples falsely classified as malicious (False positive rate). Each sample in the non-temporal model corresponds to the set of performance counter measurements in one epoch; each temporal sample spans over 4 epochs. Furthermore, to contrast the relative performance between the models in the detection of malicious samples, the area under the \emph{ROC} curve for each model can be computed and compared. This area, commonly termed as the \emph{Area Under Curve} (AUC) score, provides a quantitative measure of how well a model can distinguish between the clean and malicious samples for varying thresholds. The higher the AUC score, the better the detection performance of the model.

\subsection{Detection Performance of Models}
\label{sec:eval_auc}

\begin{figure}
  \centering
  \includegraphics[width=\columnwidth]{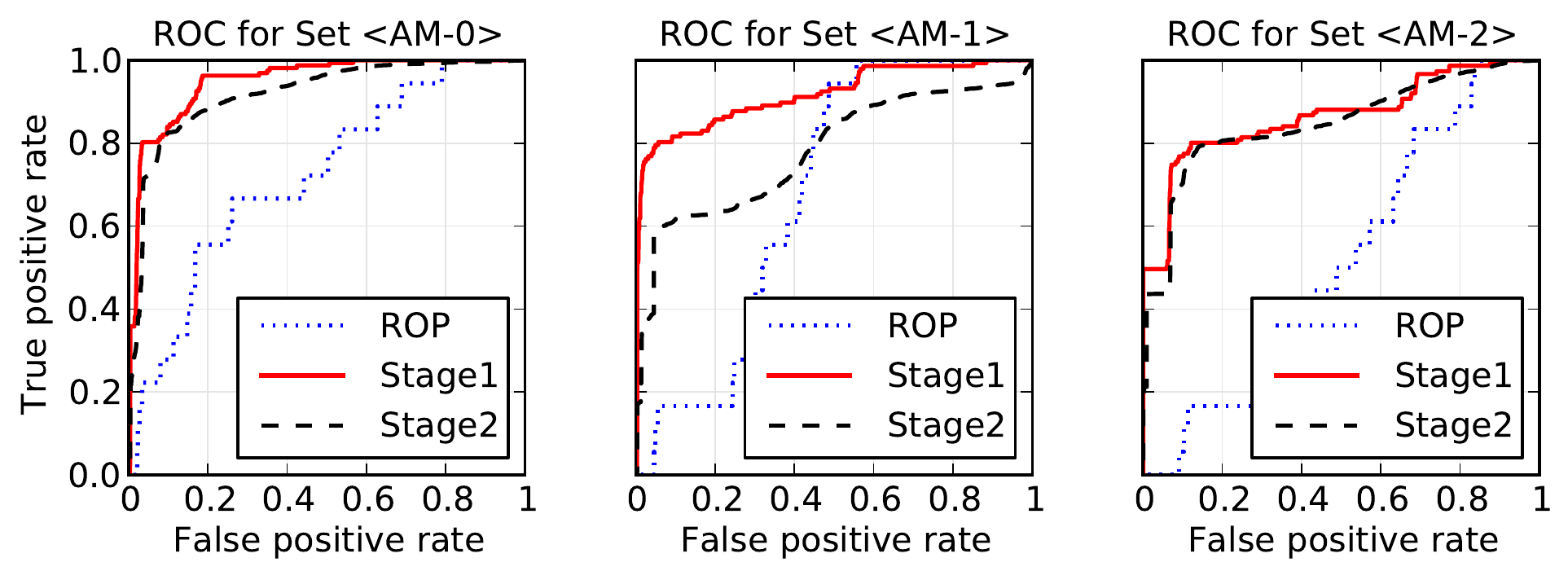}
  \caption{ROC plots for \emph{Non-Temporal} 4-feature models for IE.}
  \label{fig:roc_4d-512k}
\end{figure}

\begin{figure}
  \centering
  \includegraphics[width=\columnwidth]{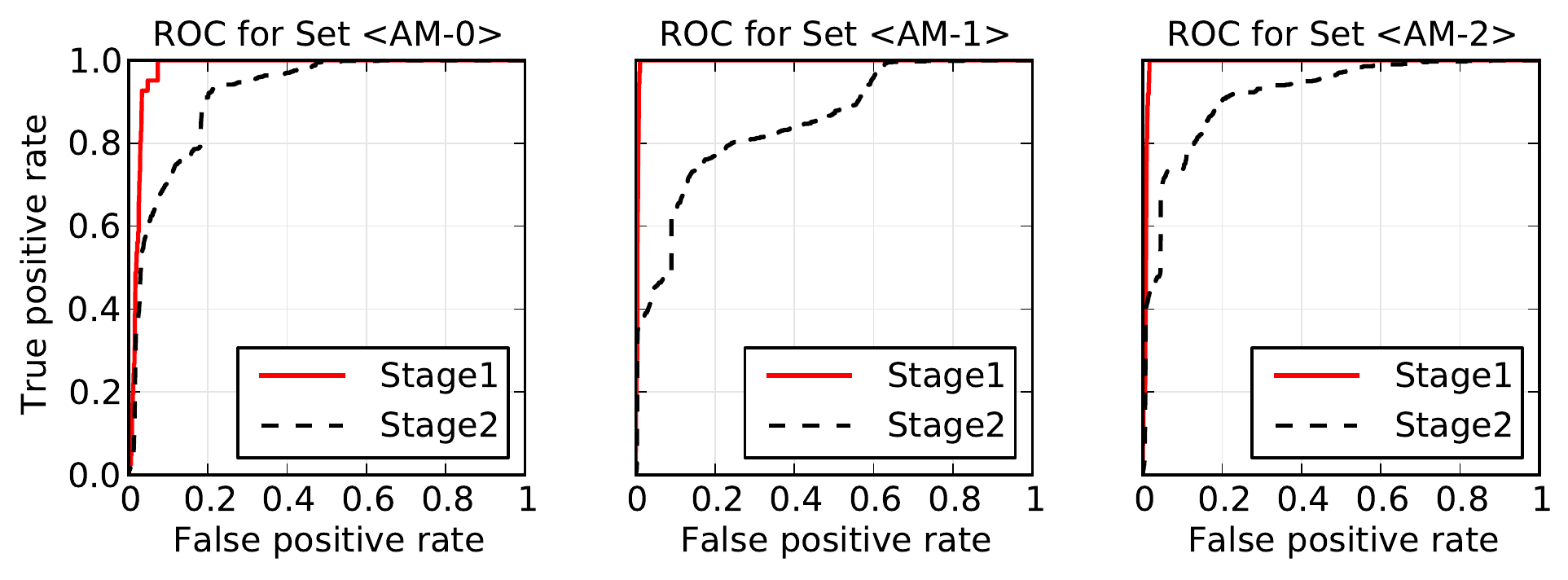}
  \caption{ROC plots for \emph{Temporal} 16-feature models for IE.}
  \label{fig:roc_16d-512k-ie}
\end{figure}

\begin{figure}
  \centering
  \includegraphics[width=\columnwidth]{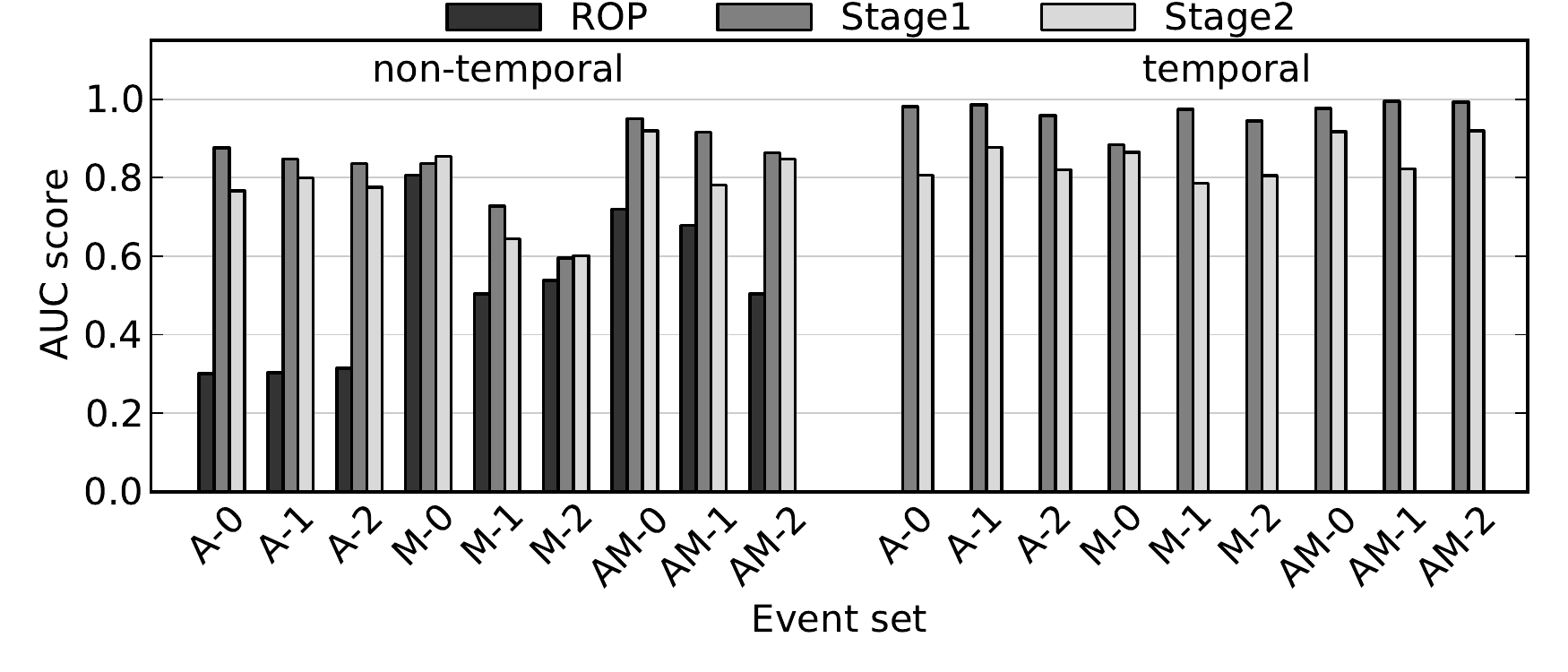}
  \caption{Detection AUC scores for different event sets using non-temporal and temporal models for IE.}
  \label{fig:combined_aucs}
\end{figure}

We first build the oc-SVM models with the training data, and then evaluate them with the testing data using the non-temporal and temporal model approaches on the nine event sets. To characterize and visualize the detection rates in terms of true and false positives over varying thresholds, we present the ROC curves of the two approaches in Figures \ref{fig:roc_4d-512k} and \ref{fig:roc_16d-512k-ie}. Due to space constraints, we only present the ROC curves for models that use both instruction-level and microarchitectural events. We also present the overall detection results in terms of AUC scores in Figure \ref{fig:combined_aucs} and highlight the key observations that affect the detection accuracy of the models below.

\subsubsection{Different Stages of Malware Exploits}
We observe that the models, in general, perform best in the detection of the \textit{Stage1} shellcode. These results suggest the \textit{Stage1} shellcode exhibits the largest deviations from the baseline architectural and microarchitectural characteristics of benign code. We achieve a best-case detection accuracy of 99.5\% for \textit{Stage1} shellcode with the \textit{AM-1} model.

On the other hand, the models show mediocre detection capabilities for the ROP shellcode. The models does not perform well in the detection of the ROP shellcode, likely because the sampling granularity at 512k instructions is too large to capture the deviations in the baseline models. While the \textit{Stage1} and \textit{Stage2} shellcode executes within several time epochs, we measured that the ROP shellcode takes 2182 instructions on average to complete execution. It ranges from as few as 134 instructions (for the Flash ROP exploit) to 6016 instructions (for the PDF ROP exploit). Since we are keeping the sampling granularity constant, the sample that contains measurements during the ROP shellcode execution will also largely consists of samples from the normal code execution.

\subsubsection{Non-Temporal vs Temporal Modeling}
We observe that the detection accuracy of the models for all event sets improves with the use of temporal information. By including more temporal information in each sample vector, we reap the benefit of magnifying any deviations that are already observable in the non-temporal approach. For one event set \textit{M-2}, this temporal approach of building the models improves the AUC score from the non-temporal one by up to 59\%.

\subsubsection{Architectural vs Microarchitectural Events}
Models built using only architectural events generally perform better than those built solely with microarchitectural events. By selecting and modeling both the most discriminative architectural and microarchitectural events together, we can achieve higher detection rates of up to an AUC score of 99.5\% for event set \textit{AM-1}.

\subsubsection{Constrained Environments}
As described in \ref{sec:m_infra}, we collect the measurements in our study from one VM and transfer the measurements to the recorder in another VM to be saved and processed. For brevity, we term this cross-remote-VM scenario where the sampling and the online classification are performed on different VMs as \textit{R-1core}.

To assess the effect on detection accuracy in the scenario where we deploy both the online classification and the measurement gathering in the same VM, we run the experiment using the model set \textit{AM-1} using two additional local-VM scenarios using 1 and 2 cores. We term these two scenarios as \textit{L-1core} and \textit{L-2core} respectively. We present the detection AUC scores for the three different scenarios in Table \ref{tbl:constrained_env}. We observe that the detection performance suffers when the online classification detector is deployed locally together with the sampling driver. This can be attributed to the possible noise that is introduced to the event counts while the online detector is continuously running and taking in the stream of samples.

\begin{table}
\resizebox{\columnwidth}{!}{%
\centering
\begin{tabular}{|c||c|c|c||c|c|} \hline
\textbf{Scenario} & \multicolumn{3}{c||}{\textbf{Non-Temporal}} & \multicolumn{2}{c|}{\textbf{Temporal}} \\ \cline{2-6}
\textbf{Label} & \textbf{ROP} & \textbf{Stage1} & \textbf{Stage2}& \textbf{Stage1} & \textbf{Stage2} \\ \hline \hline

L-1core & 0.505 & 0.895 & 0.814 & 0.918 & 0.900 \\ \hline
L-2core & 0.496 & 0.890 & 0.807 & 0.907 & 0.813 \\ \hline
R-1core & 0.678 & 0.916 & 0.781 & 0.995 & 0.823 \\ \hline

\end{tabular}
}
\caption{AUC scores for constrained scenarios using set \textit{AM-1}.}
\label{tbl:constrained_env}
\end{table}

\subsubsection{Different Sampling Granularities}
\label{sec:sampling_overhead}
While we use the sampling rate of 512k instructions for the above experiments, we also investigate the effect on detection efficacy over a range of sampling granularities.

Furthermore, while the hardware-based HPCs incur a near-zero overhead in the monitoring of the event counts, a software-only implementation of the detector still requires running programs to be interrupted periodically to sample the event counts. This inadvertently leads to a slowdown of the overall running time of programs due to this sampling overhead. To inform the deployment of a software-only implementation of such a detection paradigm, we evaluate the sampling performance overhead for different sampling rates.

\begin{figure}
  \centering
  \includegraphics[width=\columnwidth]{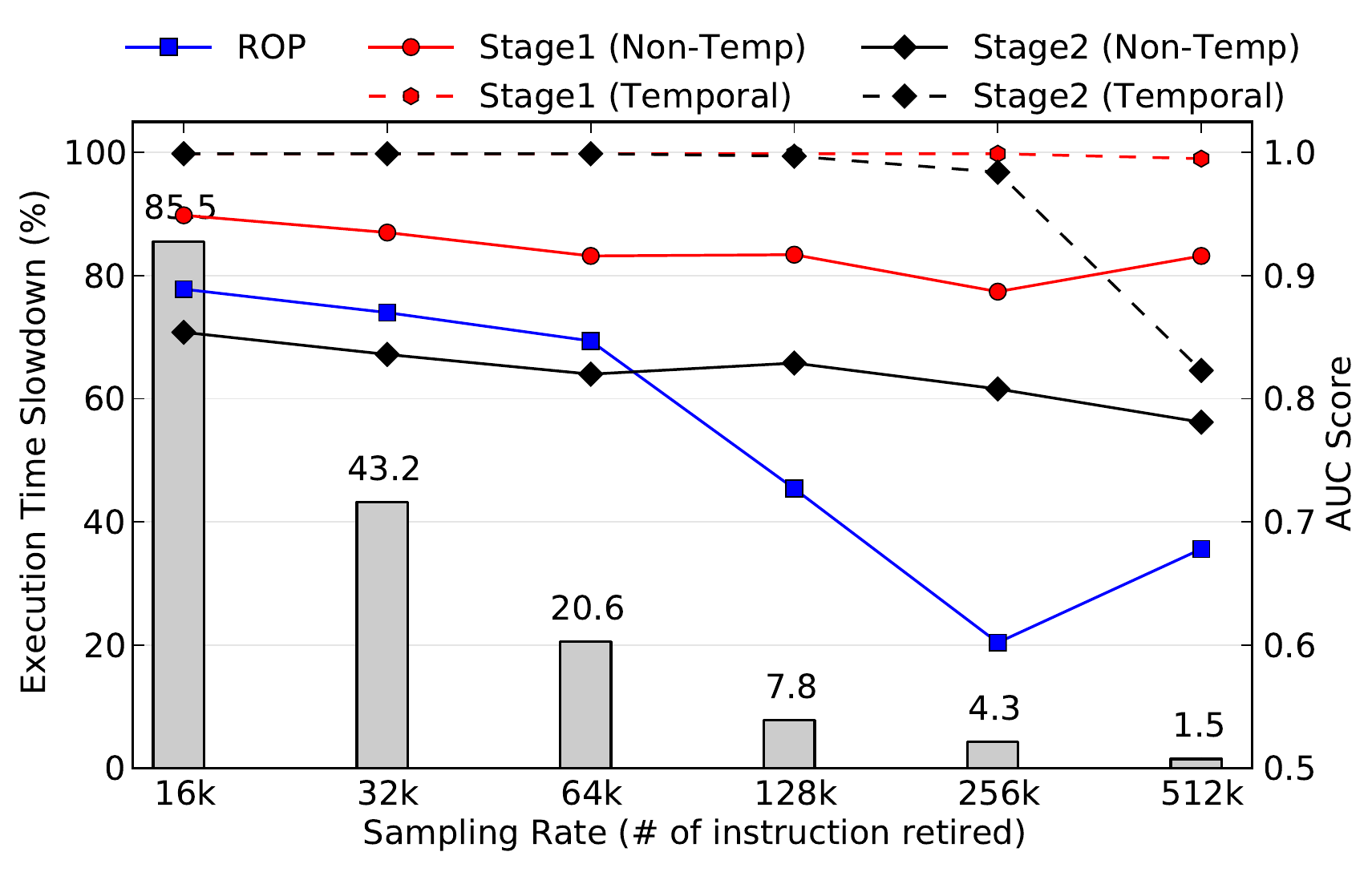}
  \caption{Trade-off between sampling overhead for different sampling rates versus detection accuracy using set \textit{AM-1}.}
  \label{fig:samp_overhead}
\end{figure}

To measure this overhead, we vary the sampling granularity and measure the slowdown in the programs from the SPEC 2006 benchmark suite \cite{Henning:2006:SCB:1186736.1186737}. We also repeated the experiments using the event set \textit{AM-1} to study the effect of sampling granularity has on the detection accuracy of the model. We plot the execution time slowdown over different sampling rates with the corresponding detection AUC scores for various malware exploit stages in Figure \ref{fig:samp_overhead}. 

We observe that the detection performance generally deteriorates with coarser-grained sampling. This is a result of the imprecise sampling technique used. For example, during the span of instructions retired in one sample, while we may label these measurements as belonging to a specific process PID, the measurements in this sample may also contain measurements belonging to other processes context-switched in and out during the span of this sample. This "noise" effect becomes more pronounced with a coarser-grained sampling rate and deteriorates the detection performance. Nonetheless, we note that the reduction in sampling overhead at coarser-grained rates far outstrips the decrease in detection performance.

\subsection{Results for Adobe PDF Reader}
Due to space constraints, we do not present the full results from our
experiments on the stand-alone Adobe PDF Reader. For brevity, we 
present the AUC detection performance of the models built with the 
event sets \textit{AM-0,1,2} in Table \ref{tbl:auc_scores_pdf}. 
Compared to the models for IE, the detection of ROP and \textit{Stage1} 
shellcode generally improves for the Adobe PDF Reader. We even achieve 
an AUC score of 0.999 with the temporal modeling of the \textit{AM-1} 
event set.  The improved performance of this detection technique for 
the PDF Reader suggests that its baseline characteristics are more 
stable given the less varied range of inputs it handles compared to IE.

\begin{table}
\resizebox{\columnwidth}{!}{%
\centering
\begin{tabular}{|c||c|c|c||c|c|} \hline
\textbf{Set} & \multicolumn{3}{c||}{\textbf{Non-Temporal}} & \multicolumn{2}{c|}{\textbf{Temporal}} \\ \cline{2-6}
\textbf{Label} & \textbf{ROP} & \textbf{Stage1} & \textbf{Stage2}& \textbf{Stage1} & \textbf{Stage2} \\ \hline \hline

AM-0 & 0.931 & 0.861 & 0.504 & 0.967 & 0.766 \\ \hline
AM-1 & 0.857 & 0.932 & 0.786 & 0.999 & 0.863 \\ \hline
AM-2 & 0.907 & 0.939 & 0.756 & 0.998 & 0.912 \\ \hline

\end{tabular}
}
\caption{AUC scores for stand-alone Adobe PDF Reader.}
\label{tbl:auc_scores_pdf}
\end{table}
\section{Analysis of Evasion Strategies}
\label{sec:evasion}
In general, anomaly-based intrusion detection approaches, such as ours, 
are susceptible to \textit{mimicry} attacks. For such an attack to evade 
detection, with sufficient information about the anomaly detection models, 
a sophisticated adversary can modify her malware into an equivalent form
that exhibits similar baseline architectural and microarchitectural 
characteristics as the normal programs. In this section, we examine the degree 
of freedom an adversary has in crafting a mimicry attack and how it impacts the 
detection efficacy of our models. 

\textbf{Adversary Assumptions} \space\space We presume the adversary has 
an exploit that she wants to execute without being detected by our models. 
We assume the adversary (1) knows all about the target 
program such as the version and OS to be run on, and (2) is able to 
gather similar HPC measurements for the targeted program to approximate 
its baseline characteristics. (3) She also knows the way the events are modeled, but \textit{not} the exact events used. We highlight three ways 
the adversary can change her attack while retaining the original 
attack semantics.

Assumption (3) is a realistic one since the modern processors allow 
hundreds of possible events to be monitored. While she may uncover the 
manner the events are modeled, it is difficult to pinpoint the exact 
subset of four events used given the numerous possible combinations of 
subsets. Furthermore, even if the entire list of events that can be 
monitored is available, there may still exist some events (such as events
monitored by the power management units) that are not publicly available.
Nonetheless, to describe the first two attacks, we optimistically assume 
that the adversary has full knowledge of all the events that are used in 
the models.

\textbf{Attack \#1: Padding} \space\space The first approach is to pad 
the original shellcode code sequences with "no-op" (no effect) 
instructions with a sufficient number
so that the events manifested by the shellcode match that of the 
baseline execution of the program. These no-op instructions should modify 
the measurements for all the events monitored, in tandem, to a range 
acceptable to the models.

The adversary needs to know the events used by the model \textit{a priori}, in 
order to exert an influence over the relevant events. We first explored 
feasibility of such a mimicry approach by analyzing the \textit{Stage1} 
shellcode under the detection model of event set \textit{AM-1}. After 
studying the true positive samples, we observe that the event 
characteristics exhibited by the shellcode are due to the unusually low 
counts of the four events modeled. As we re-craft the shellcode at the 
assembly code level to achieve the mimicry effect, we note four difficulties.

\begin{enumerate}
	\item To maintain the original semantics of the shellcode code sequences, certain registers need to be saved and subsequently restored. Be it on the stack, heap or other segments, any such operations constitute \textsc{Store} /\textsc{Load} operations, inadvertently affecting both \textsc{Store} and \textsc{Load} events.
	\item Some microarchitectural events require more than one instruction to effect a change. For example, to raise the \textsc{Misp\_Ret} event counts, code sequences of \textsc{Ret} instructions need to be crafted in a specific order. Insertion of no-ops must be added in segments.
	\item We are rarely able to craft no-op instruction segments to modify each event independently. For instance, among the four events modeled in \textit{AM-1}, the no-op instruction segment can only be crafted to affect the \textsc{Store} counts independently. Some events are modified simultaneously at different degrees with the padding of crafted no-op instruction segments.
	\item Insertion position of the no-op instruction segments can be critical to achieve the desired mimicry effect. We notice the use of several loops within the shellcode. If even one no-op segment is inserted into the loops, that results in a huge artificial increase in certain event types, consequently making that code execution look more malicious than usual.
\end{enumerate}

\begin{figure}
	  \centering
	  \includegraphics[width=\columnwidth]{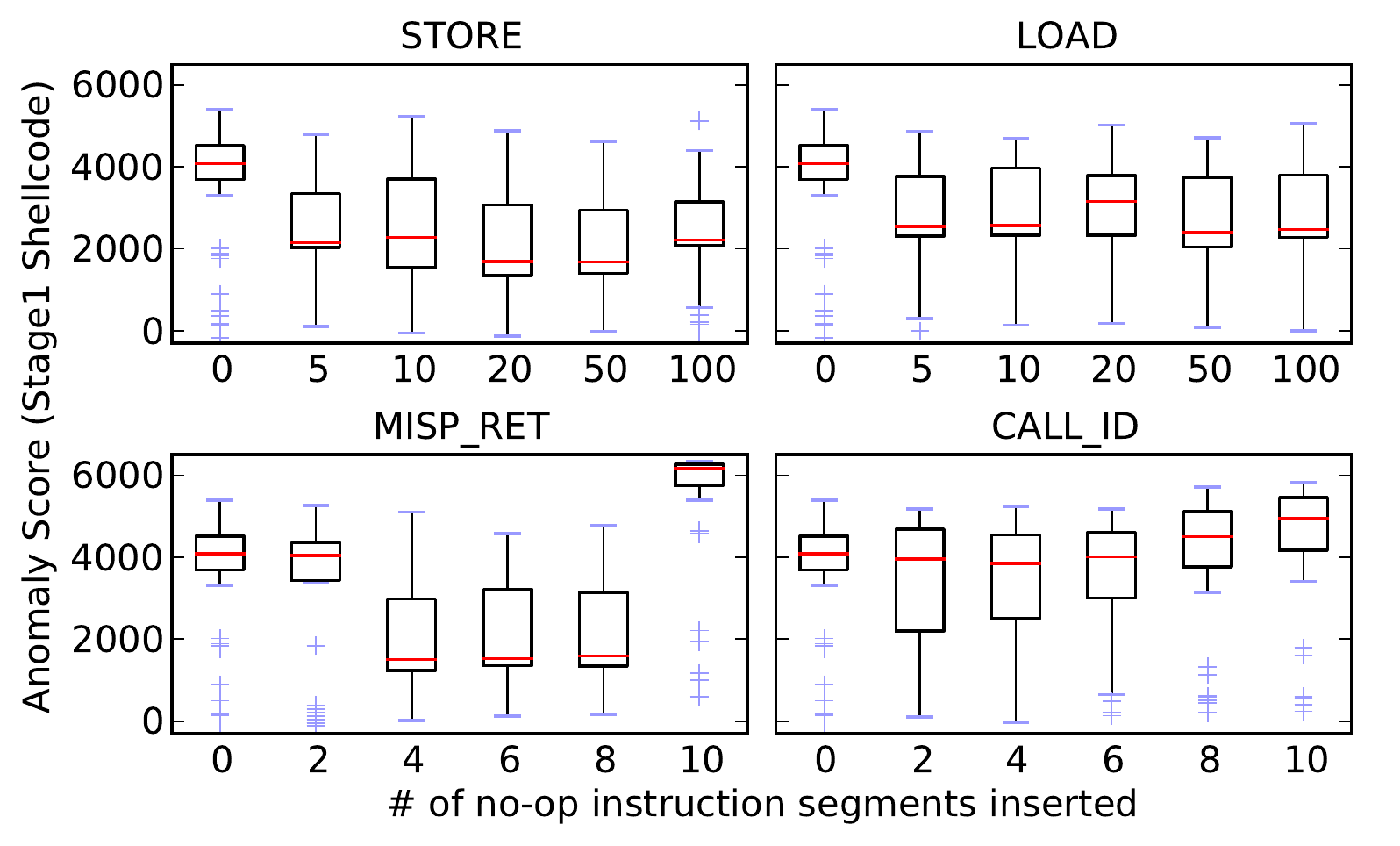}
 	 \caption{Impact of inserting no-op instruction segments on the anomaly scores of \textit{Stage1} shellcode.}
	 \label{fig:noise_boxplot}
\end{figure}

\begin{figure}
	  \centering
	  \includegraphics[width=\columnwidth]{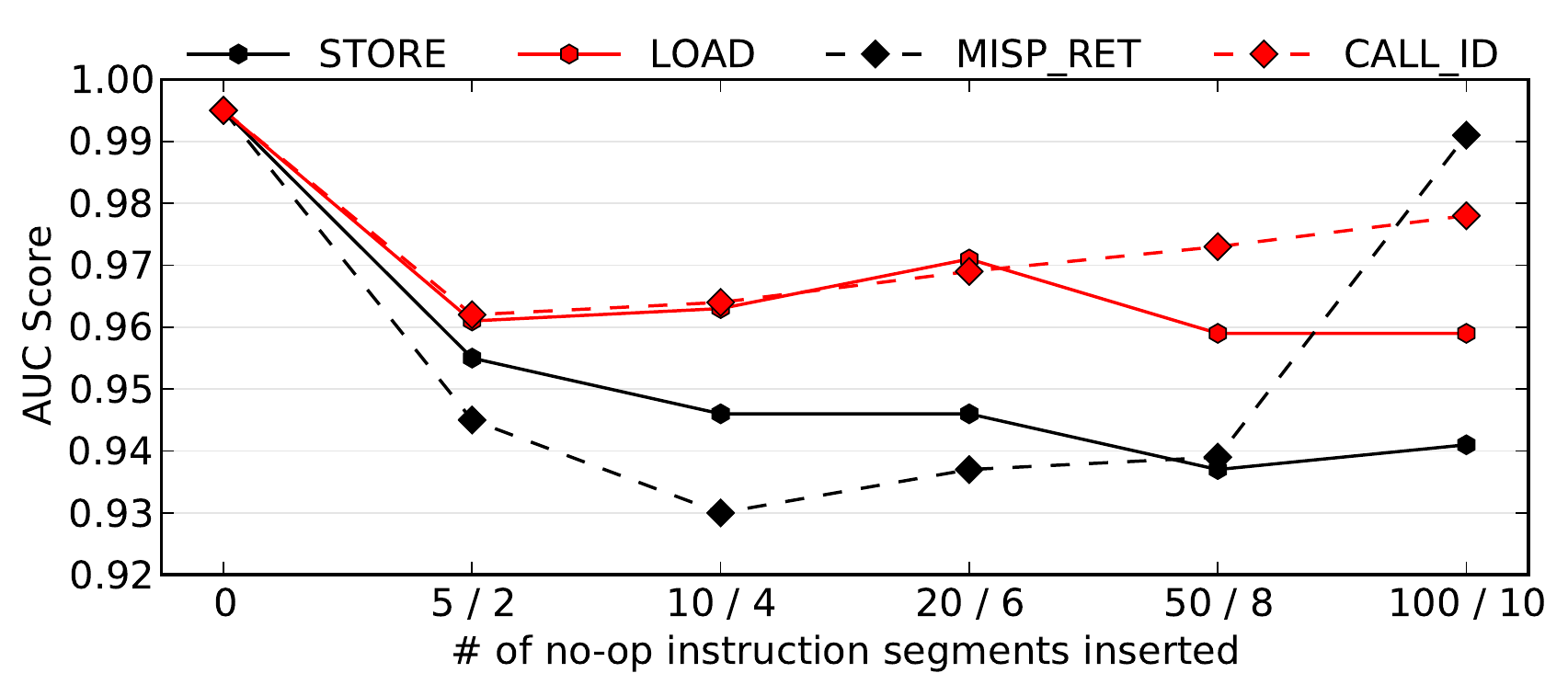}
 	 \caption{Impact of inserting no-op instruction segments on the detection performance of \textit{Stage1} shellcode.}
	 \label{fig:noise_auc}
\end{figure}

Next, we examine the impact of such mimicry efforts on the detection performance. We pad the \textit{Stage1} shellcode at random positions (avoiding the loops) with increasing number of each crafted no-op instruction segment and repeated the detection experiments. In Figure \ref{fig:noise_boxplot}, we plot the box-and-whisker plots of the anomaly scores observed from the samples with varying numbers of injected no-op code. In general, the anomaly scores become less anomalous with the padding, until after a tipping point where inserting too many no-ops reverses mimicry effect. In the same vein, we observe in Figure \ref{fig:noise_auc} that the detection AUC scores decrease as the samples appear more normal. For the worst case, the detection performance suffers by up to 6.5\% just by inserting \textit{only} the \textsc{Call\_Id} no-ops. We did not study combining the no-ops for different events, but we believe it should deteriorate the detection performance further.

\textbf{Attack \#2: Substitution} \space\space Instead of padding no-ops into original attack code sequences, the adversary can replace her code sequences with equivalent variants using code obfuscation techniques, common in  metamorphic malware \cite{borello:2008}. Like the former attack, this also requires that the she knows the events used by the models a priori.

To conduct this attack, she must first craft or generate equivalent code variants of code sequences in her exploits, and profile the event characteristics of each variant. Then she can adopt a greedy strategy by iteratively substituting parts of his attack code with the equivalent variants, measuring the HPC events of the shellcode and ditching those variants that exhibit characteristics not acceptable to the models.

However, while this greedy approach will eventually terminate, it warrants further examination as to whether the resulting shellcode modifications suffice to evade the models. We argue that this kind of shellcode re-design is hard and will substantially raise the bar for exploit writers.

\textbf{Attack \#3: Grafting} \space\space 
This attack requires either (1) inserting benign code from the 
target program directly into the exploit code, or (2) co-scheduling 
the exploit shellcode by calling benign functions (but with no-op 
effects) within the exploit code. In a sense, this attack 
\textit{grafts} its malicious code execution with the benign ones 
within the target program, thus relieving the need for the knowledge of 
the events that are modeled. If done correctly, it can exhibit very similar 
characteristics as the benign code it grafts itself to. As such this represents 
the most powerful attack against our detection approach.

While we acknowledge that we did not craft this form of attack in our 
study, we believe that it is extremely challenging to craft
such a grafting attack due to the operational constraints on the 
exploit and shellcode, described in~\ref{sec:background}. 
First, inserting sufficient benign code into 
the shellcode may exceed the vulnerability-specific size limits and
cause the exploit to fail. Second, to use benign functions for the grafting 
attacks, these  functions have to be carefully identified and inserted so that 
they execute sufficiently to mimic the normal program behavior and yet not 
interfere with the execution of the original shellcode. Third, the execution of 
grafted benign code must not unduly increase the execution time of the entire 
exploit.

\subsection{Defenses}

Unlike past anomaly-based detection systems that detect deviations based on the syntactic/semantic structure and code behavior of the malware shellcode, our approach focuses on the architectural and microarchitectural side-effects manifested through the code execution of the malware shellcode. While the adversary has complete freedom in crafting her attack instruction sequences to evade the former systems, she cannot directly modify the events exhibited by her attack code to evade our detection approach. To conduct a mimicry attack here, she has to carefully "massage" his attack code to manifest a combination of event behaviors that are accepted as benign/normal under our models. This second-order degree of control over the event characteristics of the shellcode adds difficulty to the adversary's evasion efforts. On top of this, we discuss further potential defense strategies to mitigate the impact of the mimicry attacks.

\textbf{Randomization} \space\space Introducing secret randomizations into the models has been used to strengthen robustness against mimicry attacks in anomaly-based detection systems \cite{bruschi2007efficient, wang2006anagram}. In our context, we can randomize the events used in the models by training multiple models using different subsets of the shortlisted events. We can also randomize the choice of model to utilize over time. Another degree of randomization is to change the number of consecutive samples to use for each sample for the temporal models. In this manner, the attacker does not know which model is used during the execution of his attack shellcode. For her exploit to be portable and functional a wide range of targets, she has to modify her shellcode using the no-op padding and instruction substitution mimicry attacks for a wider range of events (and not just the current four events).

To obtain a sense of the diversity introduced with this approach, we assume we have 10 different events as the pool of events we could select for the models, and that we can vary the number of consecutive samples from a range of 3 to 6. With these two degrees of freedom, the number of possible different models that can be constructed is ${10 \choose 4}\cdot4 = 840$. The number of possibilities increases substantially if we have more events in our pool. Increasing the pool of events from 10 to 20 will then result in ${20 \choose 4}\cdot4 = 19380$, a 23-fold increase.

\textbf{Multiplexing} \space\space At the cost of higher sampling overhead, we can choose to sample at a finer sampling granularity and measure more events (instead of the current four) by multiplexing the monitoring as follows. For example, we can approximate the simultaneous monitoring of 8 events across two time epochs by monitoring 4 events in one and another 4 in the other. This affords more dimensionality to the input vectors we use in the models, increasing the efforts needed by the adversary to make all the increased number of monitored event measurements look non-anomalous.

\textbf{Defense-in-depth} \space\space Consider a defense-in-depth approach, where this malware anomaly detector using HPC manifestations is deployed with existing anomaly-based detectors monitoring for other features of the malware, such as its syntactic and semantic structure \cite{krugel2002service, wang2006anagram, kong2012sa3, mahoney2003network} and its execution behavior at system-call level \cite{hofmeyr1998intrusion, somayaji2000automated, forrest1996sense, marceau2001characterizing, sekar2001fast} and function level \cite{peisert2007analysis}. In such a setting, in order for a successful attack, an adversary is then forced to shape her attack code to conform to normalcy for each anomaly detection model. An open area of research remains in quantifying this multiplicative level of security afforded by the \textit{combined} use of these HPC models with existing defenses -- i.e. examining the difficulty in shaping the malware shellcode to evade statistical and behavioral anomaly detection systems, while at the same time not exhibiting any anomalous HPC event characteristics during execution.

\section{Architectural Enhancements \\for Detection}
\label{sec:arch}

Performance counters are typically used for low-level performance
analysis and tuning, and for program characterization. In this
section we suggest some simple modifications to extend their benefits
for detecting malware based on anomalies.

\textbf{More performance counters} \space\space Our experiments show that
adding events can help better distinguish between benign and malicious
code execution. Expanding the set of performance counters that can
be monitored concurrently can potentially increase detection fidelity.
Cheap hardware mechanisms to observe instruction and data working
set changes, and basic-block level execution frequencies can improve
malware detection accuracies further.

\textbf{Interrupt-less periodic access} \space\space Currently reading
performance counters requires the host process to be interrupted.
This leads in expensive interrupt-handling cost and undue sampling
overhead to the programs. If the performance monitoring units are
re-designed with the capability to store performance counter
measurements periodically to a designated memory region without
generating interrupts, accessing the samples from this region
directly will eliminate the sampling overhead. Most importantly,
this allows for monitoring at finer granularities to reduce the
"noise" effect described in \ref{sec:sampling_overhead}, and leaves
greater scope for better detection.

\textbf{Custom Accelerators} \space\space In our work we sample at a very
coarse granularity of 512K instructions. Results show that finer
granularity sampling can improve detection accuracies. Currently
the detector is implemented in software, but at much finer
granularities, to keep up with increased data volumes, hardware
implementations will likely be necessary and certainly be more
energy-efficient compared to software implementations.

\textbf{Secret Events} \space\space In this work we have used publicly available
performance counters for detecting malware. The malware detector
can be built just as well with non-public microarchitectural events.
Keeping the events secret increases the difficulty of the attacker
to conduct evasion attacks.  This model is very similar to how
on-chip power controllers operate in modern processors.  In the
latest Intel and AMD processors, an on-chip microcontroller receives
activity factors from various blocks on the chip and uses this
information to make power management
decisions.
Neither the units providing activity factors or the logic/algorithm
for making power management decisions are public information, and
has been hard to reverse engineer.  Further the power management
algorithm is not directly accessible to software but during emergencies
an exception is delivered to the software. A similar model can be
used to build malware detectors.

\section{Related Work}
\label{sec:related}

The use of low-level hardware features for malware detection instead
of software features is a recent development.  Demme \etal demonstrates
the feasibility of misuse-based detection of Android malware programs
using microarchitectural features \cite{Demme:2013:FOM:2485922.2485970}.
While they model microarchitectural signatures of malware programs,
we build baseline microarchitectural models of programs we are
protecting and detect deviations caused by a potentially wider range
of malware (even ones that are previously unobserved). Another key
distinction is that we are detecting malware shellcode execution
of an exploit within the context of the victim program during the
act of exploitation; they target Android malware as whole programs.
After infiltrating the system via an exploit, the malware can be
made stealthier by installing into peripherals, or by infecting
other benign programs. Stewin \etal propose detecting the former
by identifying additional memory bus accesses made by the malware
\cite{stewin2013primitive}. Malone \etal examine detecting the
latter form of malicious static and dynamic program modification
by modeling the architectural characteristics
of benign programs (and excluding the use of microarchitectural events) using linear regression models
\cite{Malone:2011:HPC:2046582.2046596}. Another line of research
demonstrates that malware can be detected using side-channel power
perturbations they induce in medical embedded devices
\cite{clarkwattsupdoc}, software-defined radios
\cite{gonzalez2010detecting} and even ubiquitous mobile phones
\cite{kim2008detecting}. However, Hoffman \etal have shown that the
use of such power consumption models can be very susceptible to
noise, especially in a device with such widely varied use as the
modern smartphone \cite{hoffmann2013mobile}.

Besides the HPCs, several works have leveraged other hardware facilities on modern processors to monitor branch addresses efficiently to thwart classes of exploitation techniques. kBouncer uses the Last Branch Recording (LBR) facility to monitor for runtime behavior of indirect branch instructions during the invocation of Windows API for the prevention of ROP exploits \cite{pappastransparent}. To enforce control flow integrity, CFIMon \cite{YubinXia:2012:CDV:2354410.2355130} and Eunomia \cite{DBLP:conf/apsys/YuanXCZ11} leverage the Branch Trace Store (BTS) to obtain branch source and target addresses to check for unseen pairs from a pre-identified database of legitimate branch pairs. Unlike our approach to detecting malware, these works are designed to prevent exploitation in the first place, and are orthorgonal to our anomaly detection approach.

\section{Conclusions}
\label{sec:conclusion}

In this work, we introduce a new class of anomaly-based detectors
that use lower-level hardware features to detect malware exploits.
Unlike previously proposed signature-based detectors, they can detect
novel, unseen malware.  Thus they can be used in concert with
previously proposed signature-based predictors to better security.

A whole host of advances have enabled the creation of hardware-level anomaly-based
malware detectors. First, until very recently (2012), performance
counters were not virtualized and it was very difficult to obtain
performance data on infected production hardware systems. Second,
ironically, the availability of toolkits to create malware also
enables us to catch malware better. Like attackers we are able to
use toolkits to reproduce real-world exploitations and test detection
strategies. Finally, due to advances in computational capacity it
has become possible to run sophisticated machine learning algorithms
to perform detections in real-time.

While our experiments reveal that malware exploitation can be
detected with high accuracy, we also identify opportunities for
further advancement. We posit that the detection accuracies can be
improved with more microarchitectural observability in form of
additional counters. While we used fairly sophisticated ML algorithms,
more accurate and faster algorithms are also desirable.  Especially,
non-linear classifiers such as neural networks may be able learn
more about the structure of hidden data to improve detection
accuracies. Finally, while our detector is fairly robust to evasion
attacks, the resilience can be improved with a little additional
hardware support.

This work represents a promising advance in the field of hardware
malware detection and more broadly how computer architecture research
is conducted.  Traditionally computer architects have aimed to
manually learn program characteristics to exploit common behaviors
for performance. The effort is made difficult due to program diversity
and growth.  In this work we show program features can be learned
in an automated way. The adoption of these techniques may enable
more performant and secure machines in the future.

\bibliographystyle{abbrv}
\bibliography{references}

\end{document}